\documentclass[twocolumn,english,prl,floatfix,showpacs]{revtex4-1}
\usepackage[T1]{fontenc}
\usepackage[latin9]{inputenc}
\usepackage{amsmath}
\usepackage{amssymb}
\usepackage{graphicx}

\makeatletter

\providecommand{\tabularnewline}{\\}

\@ifundefined{textcolor}{}
{%
 \definecolor{BLACK}{gray}{0}
 \definecolor{WHITE}{gray}{1}
 \definecolor{RED}{rgb}{1,0,0}
 \definecolor{GREEN}{rgb}{0,1,0}
 \definecolor{BLUE}{rgb}{0,0,1}
 \definecolor{CYAN}{cmyk}{1,0,0,0}
 \definecolor{MAGENTA}{cmyk}{0,1,0,0}
 \definecolor{YELLOW}{cmyk}{0,0,1,0}
}

\graphicspath{ {./figures/} }

\usepackage{babel}

\makeatother

\usepackage{babel}
\begin{document}

\title{On the Relation between Random Walks and Quantum Walks }

\author{Stefan Boettcher$^{1}$, Stefan Falkner$^{1}$, and Renato Portugal$^{2}$}

\affiliation{$^{1}$Department of Physics, Emory University, Atlanta, GA 30322;
USA\\
 $^{2}$ Laboratório Nacional de Computação Cient{í}fica, Petrópolis,
RJ 25651-075; Brazil}
\begin{abstract}
Based on studies on four specific networks, we conjecture a general
relation between the walk dimensions $d_{w}$ of discrete-time random
walks and quantum walks with the (self-inverse) Grover coin. In each
case, we find that $d_{w}$ of the quantum walk takes on exactly half
the value found for the classical random walk on the same geometry.
Since walks on homogeneous lattices satisfy this relation trivially,
our results for heterogeneous networks suggests that such a relation
holds irrespective of whether translational invariance is maintained
or not. To develop our results, we extend the renormalization group
analysis (RG) of the stochastic master equation to one with a unitary
propagator. As in the classical case, the solution $\rho(x,t)$ in
space and time of this quantum walk equation exhibits a scaling collapse
for a variable $x^{d_{w}}/t$ in the weak limit, which defines $d_{w}$
and illuminates fundamental aspects of the walk dynamics, e.g., its
mean-square displacement. We confirm the collapse for $\rho(x,t)$
in each case with extensive numerical simulation. The exact values
for $d_{w}$ in themselves demonstrate that RG is a powerful complementary
approach to study the asymptotics of quantum walks that weak-limit
theorems have not been able to access, such as for systems lacking
translational symmetries beyond simple trees. 
\end{abstract}


\maketitle

\section{Introduction\label{sec:Introduction}}

Like random walks, quantum walks are rapidly gaining a central role
in describing a considerable range of phenomena, from experiments
in quantum transport~\cite{Perets08,Weitenberg11,Sansoni12,Crespi13}
to universal models of quantum computing~\cite{Childs09,Childs13}.
Quantum walks are the ``engine'' that drives quantum search algorithms~\cite{Gro97a},
with the prospect of a quadratic speed-up over classical search algorithms.
Yet, despite considerable efforts, our understanding of quantum walks
still lags behind that of random walks~\cite{Shlesinger84,Havlin87,Weiss94,Hughes96},
as they exhibit a much broader spectrum of behaviors awaiting categorization
and context, even for simple lattices~\cite{AAKV01,ambainis_2001a,Bach2004562,Childs_2002a,Childs04,Konno08,magniez_2009a,Shikano10,VA12,PortugalBook}.

For random walks, the probability density $\rho\left(\vec{x},t\right)$
to detect a walk at time $t$ at site $\vec{x}$, a distance $x=\left|\vec{x}\right|$
from its origin, obeys the scaling collapse~\cite{Havlin87}, 
\begin{equation}
\rho\left(\vec{x},t\right)\sim t^{-\frac{d_{f}}{d_{w}}}f\left(x/t^{\frac{1}{d_{w}}}\right),\label{eq:collapse}
\end{equation}
with the scaling variable $x/t^{1/d_{w}}$, where $d_{f}$ is the
(possibly fractal) dimension of the network. On a translationally
invariant lattice in any spatial dimension $d(=d_{f})$, it is easy
to show that the walk is always purely ``diffusive'', $d_{w}=2$,
with a Gaussian scaling function $f$, which is the content of many
classic textbooks on random walks and diffusion \cite{Feller66I,Weiss94}.
The scaling in Eq.~(\ref{eq:collapse}) still holds when translational
invariance is broken in certain ways or the network is fractal (i.e.,
$d_{f}$ is non-integer). However, anomalous diffusion with $d_{w}\not=2$
may arise in many transport processes~\cite{Havlin87,Bouchaud90,Redner01}.

For quantum walks, the only known value for a finite walk dimension
is that for ordinary lattices~\cite{grimmett_2004a}, where Eq.~(\ref{eq:collapse})
generically holds with $d_{w}=1$, indicating a ``ballistic'' spreading
of the quantum walk from its origin. This value has been obtained
for various versions of one and higher-dimensional quantum walks,
for instance, with so-called weak-limit theorems \cite{konno_2003a,grimmett_2004a,Segawa06,Konno08,VA12}.
The RG method we have introduced recently \cite{QWNComms13} provides
an alternative approach, expanding the analytic tools to understand
quantum walks, since it works for networks that lack translational
symmetries. While still short of the mathematical rigor of existing
limit theorems, RG provides principally exact results in terms of
the asymptotic scaling variable $x/t^{1/d_{w}}$ (or pseudo-velocity
\cite{Konno05}) whose existence allows to collapse all data for the
probability density $\rho\left(\vec{x},t\right)$, aside from oscillatory
contributions (``weak limit''), as in Eq. (\ref{eq:collapse}). 

Here, we propose a relation bridging between random and quantum walks
that elucidates their scaling properties at long times and distances
on arbitrary networks, which is intimately linked to the dynamics
of their spread as well as their algorithmic performance~\cite{Ambainis07,childs_2003b}.
We find that the walk dimension $d_{w}$ for a discrete-time quantum
walk with a Grover coin is half of that for the corresponding random
walk, 
\begin{equation}
d_{w}^{QW}=\frac{1}{2}\, d_{w}^{RW}.\label{eq:dQW=00003DdRW/2}
\end{equation}
Abstracting from four specific examples used in this paper, this relation
might be rather general, and we show that it holds even if the walks
are anomalous and the geometry lacks translational symmetry. A similar
relation has been obtained for the return probability of a continuous-time
quantum walk \cite{BM_Report}, where it is traced to the generic
long-time dominance of the ground-state eigenvalue and the fact that
$\rho$ is based on the modulo-square of the site-amplitude, instead
of linearly in the random walk case. However, such a simple connection
is not obvious here, as Eq. (\ref{eq:dQW=00003DdRW/2}) is strongly
coin-dependent.

This ability to explore a given geometry that much faster than diffusion
is essential for the effectiveness of quantum search algorithms~\cite{Ambainis07,childs_2003b}.
While this value satisfies Eq.~(\ref{eq:dQW=00003DdRW/2}), it does
little to justify it. {[}None of the existing theories, for instance,
can distinguish Eq. (\ref{eq:dQW=00003DdRW/2}) from, say, $d_{w}^{QW}=d_{w}^{RW}-1$.{]}
The simplicity and robustness of the value of $d_{w}$ is surprising,
even on a simple line, $d=1$. We can picture $\rho\left(\vec{x},t\right)$
as resulting from the superposition of all paths that lead from the
origin $\vec{x}_{0}=0$ to site $\vec{x}$ in $t$ steps, weighted
by the probability of each path. Classically, each path merely receives
a factor $\frac{1}{2}$ for the probability to branch left or right
at every step (in the simplest case). Then, all paths have the \emph{same}
weight $2^{-t}$ and $\rho\left(\vec{x},t\right)$ becomes distinguished
only by the \emph{number} of path that can reach $\vec{x}$, with
its variance after $t$ steps, $\left\langle \vec{x}^{2}\right\rangle \sim t$,
providing $d_{w}=2$. For the widely used description of a discrete-time
quantum walk~\cite{ambainis_2001a}, $\rho\left(\vec{x},t\right)$
becomes the modulo-squared of the weighted sum over the very same
paths. At any branch, each path receives a \emph{different} complex
factor to its weight. It is then the subtle superposition of these
complex weights, and their interference in the square-modulus, that
determines the spread of $\rho\left(\vec{x},t\right)$. Although quantum
walks may possess extra internal degrees of freedom, asymptotically
they invariably result in $d_{w}=1$.

\begin{table}
\centering \protect\protect\protect\caption{Fractal and walk dimensions for the networks considered here. The
classical values for $d_{f}$ and $d_{w}^{RW}$ are known for DSG~\cite{Havlin87}
and HN3~\cite{SWN}, or derived here for MK. The values for $d_{w}^{QW}$
are determined with the RG. Each case satisfies Eq.~(\ref{eq:dQW=00003DdRW/2}).
We also provide the values for translational invariant hyper-cubic
lattices for reference.}
\begin{tabular}{|c||c|c|c|}
\hline 
Network  & $d_{f}$  & $d_{w}^{RW}$  & $d_{w}^{QW}$ \tabularnewline
\hline 
Lattice & $d$ & 2 & 1\tabularnewline
MK3  & $\log_{4}(7)$  & $\log_{4}(21)\approx2.196$  & $1.098079\dots$ \tabularnewline
MK4  & $\log_{4}(13)$  & $\log_{4}\left(\frac{247}{7}\right)\approx2.571$  & $1.285253\ldots$\tabularnewline
HN3  & $2$  & $\log_{2}\left(24-8\sqrt{5}\right)\approx2.612$  & $1.305758\dots$\tabularnewline
DSG  & $\log_{2}(3)$  & $\log_{2}(5)\approx2.322$  & $1.160964\dots$\tabularnewline
\hline 
\end{tabular}\label{tab:overview} 
\end{table}

\begin{figure}[b!]
\hfill{}\includegraphics{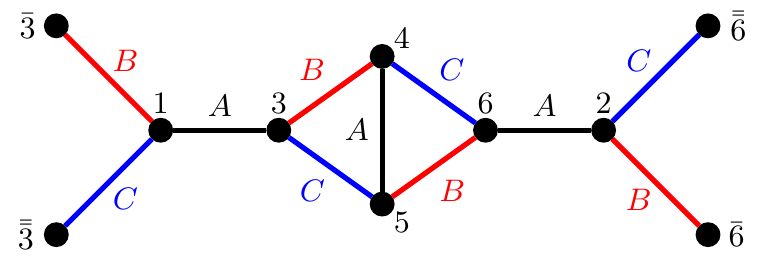} \hfill{}

\hfill{}\includegraphics{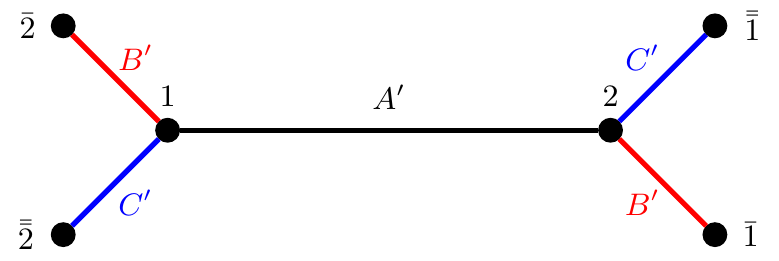} \hfill{}
\protect\protect\protect\caption{(Color online) Iterative scheme for the decimation of the Migdal-Kadanoff
network MK3. Interior sites $3,\ldots,6$ in the graph-let (top) are
algebraically eliminated, see Eq.~(\ref{eq:linear_system_mk3_1}),
and replaced by a single edge (bottom) with an effective (``renormalized'')
hopping operator $A^{\prime}$ by which the terminal site amplitudes
$1,2$ on either end of the edge shift their components between each
other. (Edges from sites $1,2$ to sites in equivalent neighboring
structures are indicated by overbars.) While renormalization is shown
for an edge of type $A$ only, types $B$ and $C$ obtain via cyclic
permutation $A\to B\to C\to A$. Constructing MK3 for simulations
proceeds by replacing every edge (bottom) by the corresponding graph-let
(top) recursively for $k$ iterations, as discussed in the Appendix.}

\label{fig:rg_illustration_mk3} 
\end{figure}

The distinct manner in which random walk and quantum walk attain their
respective probability densities $\rho\left(\vec{x},t\right)$ suggests
that a relation between their walk dimension, $d_{w}^{RW}$ and $d_{w}^{QW}$,
should be purely accidental. Any relation would be limited to a few
geometries with special constraints on quantum interference effects,
such as those imposed by translational invariance. Instead, based
on a number of diverse \emph{fractal} networks for which we have calculated
non-trivial values of $d_{w}$ for a widely used description of quantum
walks, we find the succinct relation in Eq.~(\ref{eq:dQW=00003DdRW/2})
without exception satisfied. This suggests that the common geometry
leaves a deeper imprint on the long-time behavior of both, random
and quantum walks, than might have been expected from their rather
distinct dynamics. Such insight could make quantum walk based
algorithms more predictable for networks~\cite{Paparo13}.

This paper is organized as follows: In the next section, we introduce
the formulation of the discrete-time quantum walk that we will use in the
RG analysis. In Sec. \ref{sec:Quantum-Walk-Renormalization}, we discuss
the RG procedure by example of the simplest of our networks and use
it to discuss the results for all networks, while details of the calculations
for most of those networks are provided in the Appendix. In Sec. \ref{sec:Conclusions}
we conclude discussing the implication of our results for
universality, and give an outlook on future studies.

\section{Discrete Time Quantum Walks\label{sec:Discrete-Time-Quantum}}

The dynamics for a discrete-time walk with a coin, classical or quantum,
is determined by the master-equation, 
\begin{equation}
\left|\Psi_{t+1}\right\rangle ={\cal U}\left|\Psi_{t}\right\rangle .\label{eq:MasterE}
\end{equation}
In the site-basis $\left|\vec{x}\right\rangle $ of any network, we
can describe the state of the system in terms of the site amplitudes
$\psi_{\vec{x},t}=\left\langle \vec{x}|\Psi_{t}\right\rangle $. For
a classical random walk, the probability density in Eq.~(\ref{eq:collapse})
is simply given by the site amplitude itself, $\rho(\vec{x},t)=\psi_{\vec{x},t}$,
while for the quantum walk it is $\rho(\vec{x},t)=\left|\psi_{\vec{x},t}\right|\,^{2}$.
Accordingly, the propagator ${\cal U}$ is a stochastic Bernoulli
coin for a random walk, while it must be \emph{unitary} for a quantum
walk, usually composed~as 
\begin{equation}
{\cal U}={\cal S}\left(\mathbb{I}\otimes{\cal C}\right),\label{eq:propagator}
\end{equation}
with coin ${\cal C}$ and shift ${\cal S}$. Unitarity, ${\cal U}^{\dagger}{\cal U}=\mathbb{I}$,
demands~\cite{Meyer96,Portugal14} that the coin is a unitary matrix
of rank $r>1$, such that the site amplitudes $\psi_{\vec{x},t}$
become complex $r$-dimensional vectors in ``coin''-space. For simplicity,
this quantum walk is commonly studied on networks of regular degree
$r$ for all $\vec{x}$, so that the same coin can be applied at every
site. Every step consists of a ``coin flip'', the multiplication
of $\psi_{\vec{x},t}$ with ${\cal C}$, followed by the shift ${\cal S}$
that transfers each component of ${\cal C}\cdot\psi_{\vec{x},t}$
to exactly one of the $r$ neighbors of $\vec{x}$.

\begin{figure}
\centering \includegraphics{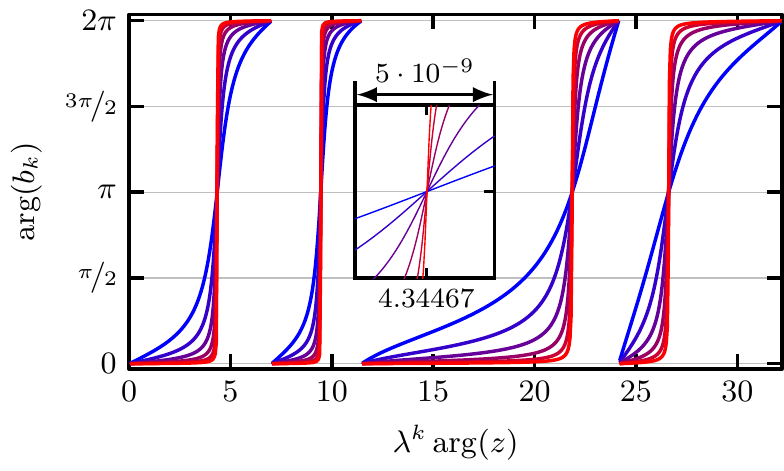} \protect\protect\protect\caption{(Color online) Scaling collapse for MK3 of the phase of $b_{k}$ in
Eq.~(\ref{eq:rg_recursions_mk3}) near the fixed point $z=1$ with
$\lambda=\sqrt{21}$. The inset shows the region around the first
intersection. In the main panel, $k=4,6,\dots,14$ while $k=50,52,\dots60$
for the inset, corresponding to a system size of MK3 with up to $N\approx7^{60}\approx10^{51}$
sites. }
\label{fig:mk3_rg_plot} 
\end{figure}

\begin{figure}
\centering \includegraphics[bb=0bp 65bp 244bp 202bp,clip,scale=0.9]{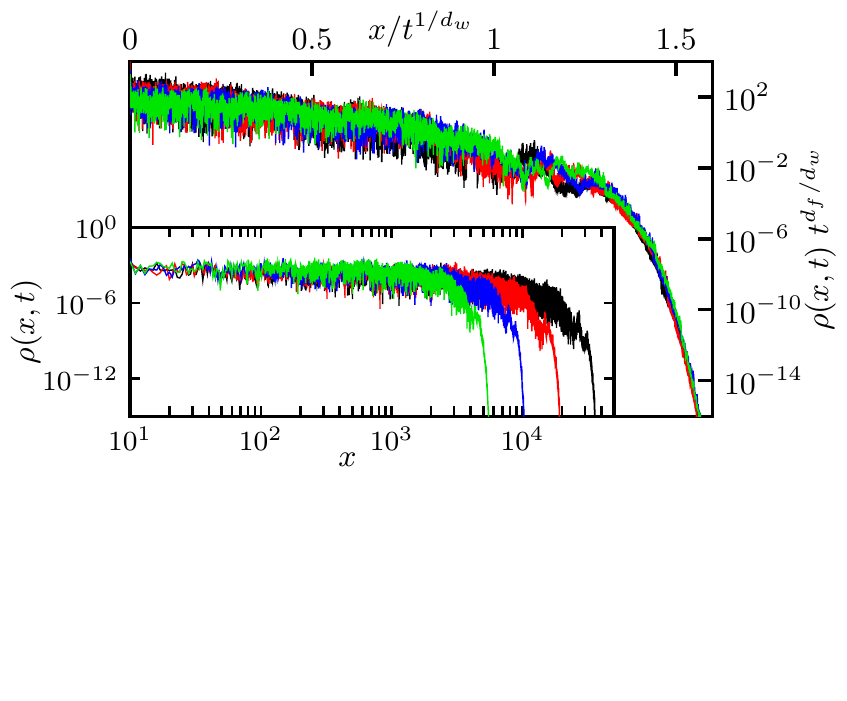}

\centering \includegraphics[bb=0bp 65bp 244bp 200bp,clip,scale=0.9]{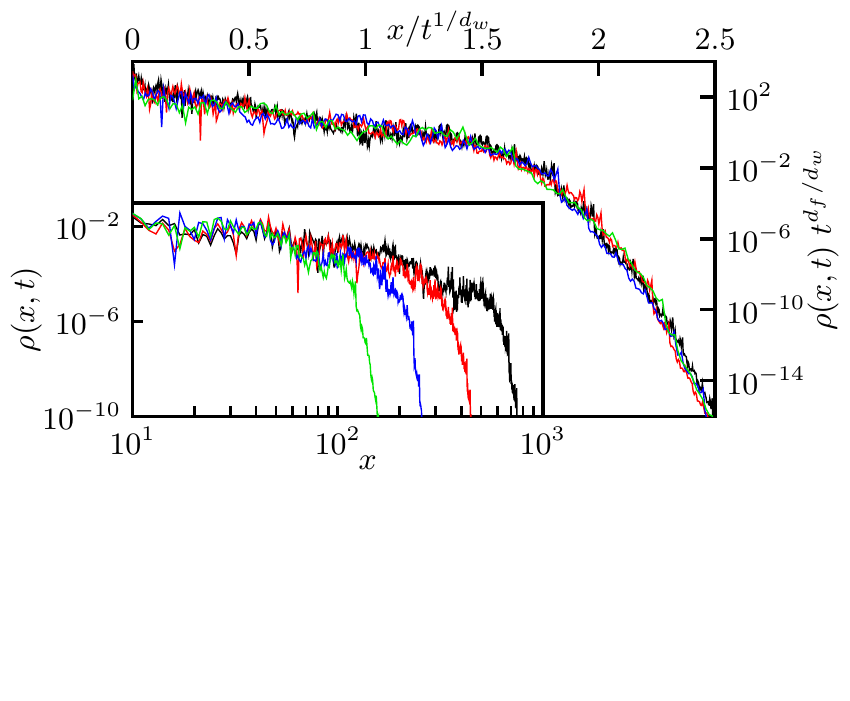}

\centering \includegraphics[bb=0bp 65bp 244bp 200bp,clip,scale=0.9]{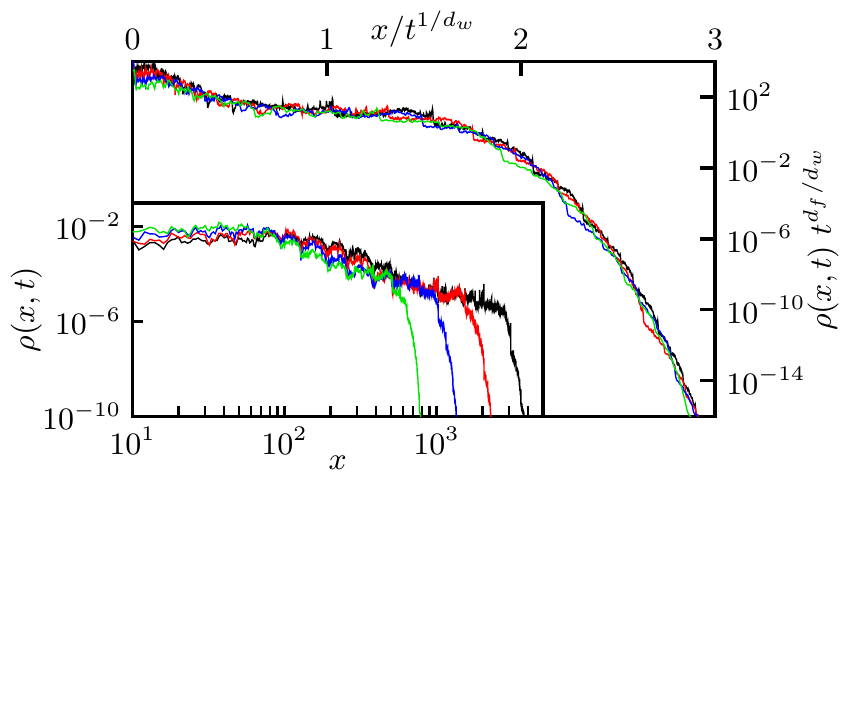}

\centering \includegraphics[bb=0bp 65bp 244bp 200bp,clip,scale=0.9]{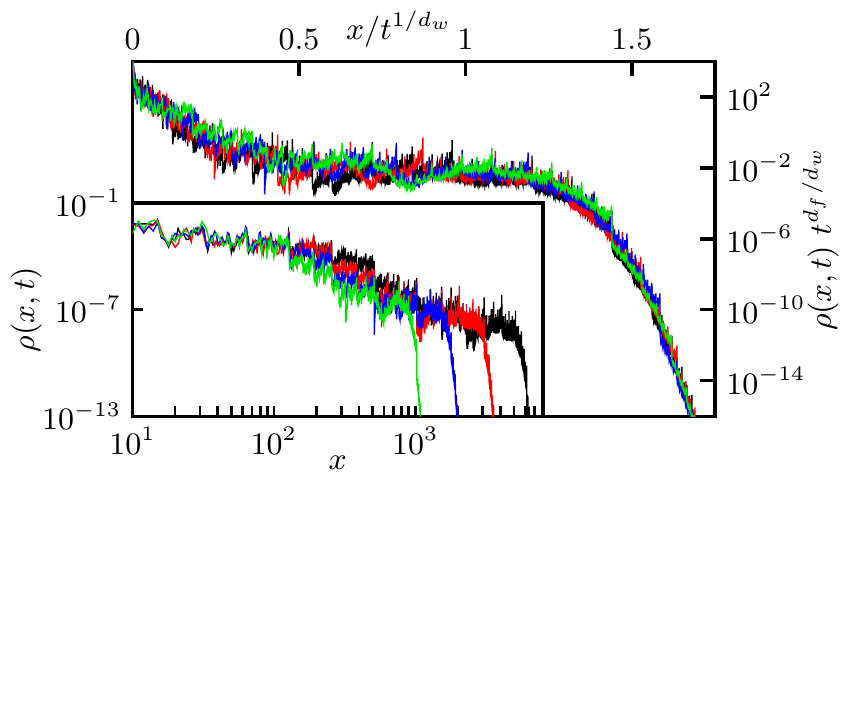}
\protect\protect\protect\caption{(Color online) Data collapse of the probability density $\rho(|\vec{x}|,t)$
according to Eq.~(\ref{eq:collapse}) with $d_{f}$ and $d_{w}$
given in Tab.~\ref{tab:overview}. The data are obtained by direct
simulations of quantum walks on the four different networks in this
study. The inset of each panel shows the raw data, from left to right
for increasing time in each case. The top panel concerns MK3 with
$N=2\cdot7^{8}\approx10^{7}$ sites at times $t=2^{j}$, $j=13,\ldots,16$.
In the main panel, the data are collapse with $d_{f}=\log_{4}(7)$
and $d_{w}^{QW}=\log_{4}(21)/2$. The $2^{{\rm nd}}$ panel concerns
MK4 with $N=2\cdot13^{6}\approx10^{7}$ sites at $t=2^{j}$,
$j=12,\ldots,15$, collapsed with $d_{f}=\log_{4}(13)$ and $d_{w}^{QW}=\log_{16}(247/7)$.
The $3^{{\rm rd}}$ panel concerns HN3 with $N=2^{24}\approx1.7\cdot10^{7}$
sites at $t=2^{j}$, $j=11,\ldots,14$, collapsed with $d_{f}=2$
and $d_{w}^{QW}=\log_{4}\left(24-8\sqrt{5}\right)$. The $4^{{\rm th}}$
panel on the bottom concerns DSG with $N=3^{15}\approx1.4\cdot10^{7}$
sites at $t=2^{j}$, $j=11,\ldots,14$, collapsed with $d_{f}=\log_{2}(3)$
and $d_{w}^{QW}=\log_{2}(5)/2$. }
\label{fig:direct_simulations} 
\end{figure}

To test Eq.~(\ref{eq:dQW=00003DdRW/2}) for nontrivial values for
$d_{w}$, we study the quantum walk on four fractal networks of degrees
$r=3$ and 4, with the widely used Grover coin~\cite{Gro97a,PortugalBook}.
Namely, we study two Migdal-Kadanoff networks~\cite{Berker79,Plischke94}
(MK3 and MK4), the dual Sierpinski gasket~\cite{Havlin87,Hughes96}
(DSG), and the Hanoi network~\cite{SWN} (HN3). These networks lack
translational invariance, but exhibit self-similarity instead. DSG
more closely resembles a $2d$ lattice, MK networks have a hierarchical
structure, while HN3 is a hyperbolic~\cite{Singh14} small-world
network. For each network, the anomalous classical result for $d_{w}$
of the random walk and their fractal dimension $d_{f}$ are easily
obtained via the renormalization group (RG) method, which is discussed
in many textbooks on statistical physics~\cite{Plischke94,Hughes96}
and on transport properties \cite{Redner01}. (We have provided a
simple primer in the context of quantum walks in \cite{Boettcher13a}.)
We describe the application of RG below for MK3; the RG for MK4, DSG
and HN3 is discussed in the Appendix. By extending RG to quantum walks~\cite{QWNComms13},
we obtain the first exact scaling exponents for quantum walks on heterogeneous
structures. All results are summarized in Table~\ref{tab:overview}.

\section{Quantum Walk RG for MK3\label{sec:Quantum-Walk-Renormalization}}

The fractal dimension~\cite{Havlin87,Hughes96} is defined via the
scaling $N\sim L^{d_{f}}$, where $N$ stands for the number of sites
that are at most $L$ hops away from a given site. For MK3, as shown
in Fig.~\ref{fig:rg_illustration_mk3}, the number of edges (and,
hence, sites) changes $7$-fold between iterations while distances
between two sites changes 4-fold, implying $d_{f}=\log_{4}(7)$.

To calculate the walk dimension with RG, we first apply the Laplace
transform~\cite{Weiss94,Hughes96,Redner01}, 
\begin{equation}
\left|\tilde{\Psi}\left(z\right)\right\rangle =\sum_{t=0}^{\infty}z^{t}\left|\Psi_{t}\right\rangle ,\label{eq:laplace_transform}
\end{equation}
to Eq.~(\ref{eq:MasterE}), providing algebraic equations with generalized
hopping operators that now depend on $z$. For instance, after any
number of iterations, MK3 entirely consist of graphlets, as depicted
in the top panel of Fig.~\ref{fig:rg_illustration_mk3}. For sites
$3,\dots,6$, it represents the \emph{linear} system of equations~\cite{PhysRevB.28.3110}
\begin{equation}
\begin{pmatrix}\tilde{\psi}_{3}\\
\tilde{\psi}_{4}\\
\tilde{\psi}_{5}\\
\tilde{\psi}_{6}
\end{pmatrix}=\begin{bmatrix}A & 0 & M & B & C & 0\\
0 & 0 & B & M & A & C\\
0 & 0 & C & A & M & B\\
0 & A & 0 & C & B & M
\end{bmatrix}\cdot\begin{pmatrix}\tilde{\psi}_{1}\\
\tilde{\psi}_{2}\\
\tilde{\psi}_{3}\\
\tilde{\psi}_{4}\\
\tilde{\psi}_{5}\\
\tilde{\psi}_{6}
\end{pmatrix}\label{eq:linear_system_mk3_1}
\end{equation}
with hopping operators $A$, $B$, $C$, and $M$, where $M$ allows
for self-interaction at each site. (In the original graph $M=0$.)
Taking advantage of self-similarity, we express one iteration of the
network in terms of the next smaller one but with ``renormalized''
values for the hopping operators. To that end, we solve for $\tilde{\psi}_{3},\dots\tilde{\psi}_{6}$
in term of $\tilde{\psi}_{1}$ and $\tilde{\psi}_{2}$ and insert
into the equations for the remaining site amplitudes, such that 
\begin{align}
\tilde{\psi}_{1,2} & =M\tilde{\psi}_{1,2}+A\tilde{\psi}_{3,6}+B\tilde{\psi}_{\overline{3},\overline{6}}+C\tilde{\psi}_{\overline{\overline{3}},\overline{\overline{6}}},\\
 & =M^{\prime}\tilde{\psi}_{1,2}+A^{\prime}\tilde{\psi}_{2,1}+B^{\prime}\tilde{\psi}_{\overline{2},\overline{1}}+C^{\prime}\tilde{\psi}_{\overline{\overline{2}},\overline{\overline{1}}},\nonumber
\end{align}
where primes indicate the renormalized hopping operators as depicted
in the bottom panel of Fig.~\ref{fig:rg_illustration_mk3}. Repetition
then relates the $k+1$ (primed) iterate to the $k$-th (unprimed)
iterate, yielding the
RG-flow~\cite{Plischke94,Redner01}
\begin{equation}
(A_{k+1},B_{k+1},C_{k+1},M_{k+1})=\mathcal{RG}\left(A_{k},B_{k},C_{k},M_{k}\right)
\label{eq:rg_equation_abstact}
\end{equation}
that characterizes the effective dynamics between domains of sites
of width $L_{k}$ and $L_{k+1}$ by renormalized hopping operators.

In case of the unbiased random walk, all the hopping operators become
simple scalars, $A=B=C=a$, and setting $M=1-b$, Eq.~(\ref{eq:rg_equation_abstact})
provides 
\begin{equation}
\begin{aligned}a_{k+1} & =\frac{2a_{k}^{4}}{b_{k}^{3}-4a_{k}^{2}b_{k}-a_{k}b_{k}^{2}},\\
b_{k+1} & =b_{k}+\frac{3a_{k}^{2}(2a_{k}-b_{k})(a_{k}+b_{k})}{b_{k}^{3}-4a_{k}^{2}b_{k}-a_{k}b_{k}^{2}},
\end{aligned}
\label{eq:rg_recursions_mk_rw}
\end{equation}
with the initial conditions $a_{0}=z/3$ and $b_{0}=1$. For $z\to1$,
the relevant fixed point (describing the infinite system, $k\to\infty$)
is $a_{\infty},b_{\infty}\to0$, i.e., the width of the domains grows as $L_k\sim4^k$ which is faster than the reach of the diffusive transport between them,  as represented
by $a_{k}$. With the scaling Ansatz $a_{k}=3^{-k}\alpha_{k}$ and
$b_{k}=3^{-k}\beta_{k}$, we resolve this boundary layer to find the
fixed point $\beta_{\infty}=3\alpha_{\infty}$ with Jacobian eigenvalue
$\lambda=21$ that relates to the rescaling of time, $T_{k+1}=\lambda T_{k}$,
by the Tauberian theorems~\cite{Weiss94,Hughes96,Redner01}. Then,
$L_{k+1}=4L_{k}$ and $T_{k}\sim L_{k}^{d_{w}}$ from Eq.~(\ref{eq:collapse}),
finally yield $d_{w}^{RW}=\log_{4}(21)$.

For the quantum walk, the hopping operators now are matrices in coin-space,
and the algebra gets more involved. Iterating the matrix-valued RG-flow
in Eq.~(\ref{eq:rg_equation_abstact}) numerically suggests that
all matrices can be parametrized with merely two scalars, most conveniently
in the form $\{A,B,C\}=\frac{a+b}{2}\left(P_{\{1,2,3\}}\cdot\mathcal{C}_{G}\right)$
and $M=\frac{a-b}{2}\left(\mathbb{I}\cdot\mathcal{C}_{G}\right)$,
where the $3x3$-matrices $\left[P_{\nu}\right]_{i,j}=\delta_{i,\nu}\delta_{\nu,j}$
(with $\sum_{\nu=1}^{3}P_{\nu}=\mathbb{I}$) facilitate the shift
of the $\nu$-th component to a neighboring site. The RG-flow closes
for 
\begin{widetext}
\begin{equation}
\begin{aligned}a_{k+1} & =\frac{-9a_{k}+5a_{k}^{3}+9b_{k}+3a_{k}b_{k}-17a_{k}^{2}b_{k}-3a_{k}^{3}b_{k}+3b_{k}^{2}+14a_{k}b_{k}^{2}-3a_{k}^{2}b_{k}^{2}-18a_{k}^{3}b_{k}^{2}}{-18-3a_{k}+14a_{k}^{2}+3a_{k}^{3}-3b_{k}-17a_{k}b_{k}+3a_{k}^{2}b_{k}+9a_{k}^{3}b_{k}+5b_{k}^{2}-9a_{k}^{2}b_{k}^{2}},\\
b_{k+1} & =\frac{-3a_{k}-a_{k}^{2}+3b_{k}+4a_{k}b_{k}-3a_{k}^{2}b_{k}-b_{k}^{2}+3a_{k}b_{k}^{2}+6a_{k}^{2}b_{k}^{2}}{6+3a_{k}-a_{k}^{2}-3b_{k}+4a_{k}b_{k}+3a_{k}^{2}b_{k}-b_{k}^{2}-3a_{k}b_{k}^{2}},
\end{aligned}
\label{eq:rg_recursions_mk3}
\end{equation}

\end{widetext}

with $a_{0}=b_{0}=z$. It can be shown that $|a_{k}|=|b_{k}|\equiv1$
for all $k$, reducing the RG parameters to just two real phases for
$a_{k},b_{k}$.

As explained in Ref.~\cite{QWNComms13}, the classical fixed-point
analysis from above fails for the quantum walk. Unitary demands that
information about $\rho(\vec{x},t)$ has to be recovered from an integral
involving $\tilde{\psi}_{\vec{x}}\left[a_{k}(z),b_{k}(z)\right]$
around the unit circle in the complex-$z$ plane. It is the \emph{scaling
collapse} of $\left\{ a,b\right\} _{k}(z)\sim f_{\{a,b\}}\left(\lambda^{k}\arg z\right)$,
and consequently of any observable function of $\tilde{\psi}_{\vec{x}}$,
over a finite support that allows to approximate $d_{w}$ recursively
with arbitrary accuracy. An illustration of the collapse for, say,
the phase of $b_{k}$ is shown in Fig.~\ref{fig:mk3_rg_plot}. Equivalent
plots can be found in the Appendix for MK4, HN3, and DSG.

To justify these RG predictions for $d_{w}$, we resort to direct
simulation of quantum walks to test Eq.~(\ref{eq:collapse}). Those
simulations cannot reach as extreme a system sizes as RG, but the
collapse of the probability density $\rho\left(x,t\right)$ over the
entire network illustrates the consistency with the RG predictions,
as shown in Fig.~\ref{fig:direct_simulations} for all four networks
considered here.

\section{Conclusions\label{sec:Conclusions}}

We have shown how to apply RG to obtain the scaling for the limit
distribution in Eq. (\ref{eq:collapse}) for discrete-time quantum
walks on several network for which RG is exact. This study demonstrates
that RG can deliver unprecedented insights into the dynamics of quantum
processes on systems that lack those symmetries familiar from lattices, hypercubes, trees, etc., such as translational invariance. While RG is limited to specific networks
such as those considered here (which may not in themselves be of technical
importance), conceptually, the accumulation of the obtained results
suggests a larger picture. Our findings hint at a deep, residual connection
between classical and quantum walks based on the geometry of the network
they share, which is surprising in light of the often dramatic quantum
interference effects that distinguish quantum walks from random walks.
The conjecture in Eq. (\ref{eq:dQW=00003DdRW/2}) is likely not a
trivial result. We have evidence for this simple relation to hold
only for the Grover coin, which has the property of being reflective,
making it is its own inverse. Other coins without that property, indeed,
lead to different asymptotic limits, as we will describe elsewhere.
This raises interesting questions regarding the range of possible
\emph{universality classes} of these results and their origin, a central
concern of RG~\cite{Plischke94} that has remained largely unexplored
for quantum walks~\cite{QWNComms13}. In turn, it is straightforward
to show that, asymptotically, random walks on these networks are independent
of the specific choices for a Bernoulli coin. However, for quantum
walks, the most general unitary coin matrix ${\cal C}$ for $r=3$
would already contain six free parameters that could impact the dynamics
in unforeseen ways, and could lead to significant means of control.

\paragraph*{Acknowledgements:}

SB and SF acknowledge financial support from the U. S. National Science
Foundation through grant DMR-1207431. SB acknowledges financial support
from CNPq through the ``Ciência sem Fronteiras'' program and thanks
LNCC for its hospitality. RP acknowledges financial support from Faperj
and CNPq.

\section*{Appendix}

While the methods presented for MK3 in the main text directly transfer
to the other networks, we shall outline the procedure for them in
more detail here. First, we consider the case of MK4, then we will
discuss HN3 and the dual Sierpinski gasket (DSG) that has been discussed
previously \cite{QWNComms13}. MK4 is similar to MK3 except that it
features a different degree for each site and, thus, establishes the
conjecture for a different, rank $r=4$ Grover coin than for the other
networks considered here, which all use the Grover coin of rank $r=3$.
We have focused on the lowest-rank coins because higher-ranked coins
generally make the algebra more complex. However, this $r=4$ result
demonstrates that the conjecture is likely robust, irrespective of the degree of sites..

\paragraph*{RG for MK4:}

\begin{figure}
\includegraphics{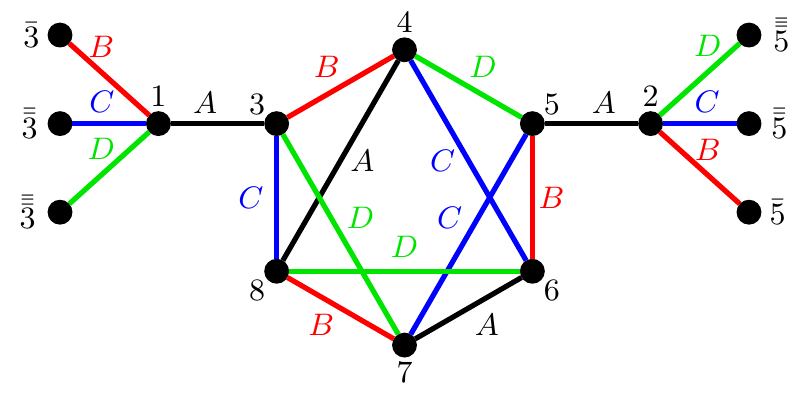} \includegraphics{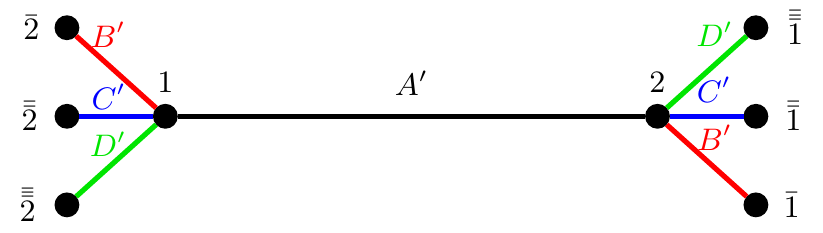}
\protect\caption{(Color online) Iteration scheme for MK4. The six interior nodes $3,\dots,8$
and all their connections (top) are replaced by direct connection
between $1$ and $2$ (bottom). The renormalized hopping parameter
$A'$ depends on all hopping matrices in the previous step. The construction
of the network can be seen as the reverse process, inserting 6 nodes
into every edge leaving the hopping parameter unchanged. The nodes
labeled with overbars represent to analogous nodes where the same
rule is applied. The scheme for $B$, $C$ and $D$ is obtained by
cyclic permutation of the shown graphlets.}
\label{fig:decimation_mk4} 
\end{figure}

MK4 follows the same idea as MK3 as every edge is replaced by multiple
nodes and edges from one generation to the next. The smallest four-regular
graph that can be consistently labeled with four different edge types
such that every node is connected to one of each kind contains 6 nodes,
see Fig.~\ref{fig:decimation_mk4}. From the graphical representation,
we can  read of the linear system for the Laplace-transformed
amplitudes on the interior nodes: 
\begin{equation}
\begin{pmatrix}\tilde{\psi}_{3}\\
\tilde{\psi}_{4}\\
\tilde{\psi}_{5}\\
\tilde{\psi}_{6}\\
\tilde{\psi}_{7}\\
\tilde{\psi}_{8}
\end{pmatrix}=\begin{bmatrix}A & 0 & M & B & 0 & 0 & D & C\\
0 & 0 & B & M & D & C & 0 & A\\
0 & A & 0 & D & M & B & C & 0\\
0 & 0 & 0 & C & B & M & A & D\\
0 & 0 & D & 0 & C & A & M & B\\
0 & 0 & C & A & 0 & D & B & M
\end{bmatrix}\cdot\begin{pmatrix}\tilde{\psi}_{1}\\
\tilde{\psi}_{2}\\
\tilde{\psi}_{3}\\
\tilde{\psi}_{4}\\
\tilde{\psi}_{5}\\
\tilde{\psi}_{6}\\
\tilde{\psi}_{7}\\
\tilde{\psi}_{8}
\end{pmatrix}\,.\label{eq:linear_system_mk4_1}
\end{equation}
Once the solution in terms of $\tilde{\psi}_{1}$ and $\tilde{\psi}_{2}$
is found, we can plug it into the equations for, say, $\tilde{\psi}_{1}$
\begin{equation}
\tilde{\psi}_{1}=A\tilde{\psi}_{3}+B\tilde{\psi}_{\bar{3}}+C\tilde{\psi}_{\bar{\bar{3}}}+D\tilde{\psi}_{\bar{\bar{\bar{3}}}}\label{eq:linear_system_mk4_2}
\end{equation}
to find the renormalized system 
\begin{equation}
\tilde{\psi}_{1}=A'\tilde{\psi}_{2}+B'\tilde{\psi}_{\bar{2}}+C'\tilde{\psi}_{\bar{\bar{2}}}+D'\tilde{\psi}_{\bar{\bar{\bar{2}}}}\label{eq:renormalized_system_mk4}
\end{equation}
By studying the first few iterations, we choose 
\begin{equation}
\begin{aligned}A_{k} & =\frac{a+b}{2}(P_{1}\cdot\mathcal{C}_{G})\,, & B_{k} & =\frac{a+b}{2}(P_{2}\cdot\mathcal{C}_{G})\,,\\
C_{k} & =\frac{a+b}{2}(P_{3}\cdot\mathcal{C}_{G})\,, & D_{k} & =\frac{a+b}{2}(P_{4}\cdot\mathcal{C}_{G})\,,\\
M_{k} & =\frac{a-b}{2}\cdot(\mathbb{I}\cdot\mathcal{C}_{G})
\end{aligned}
\label{eq:ansatz_mk4}
\end{equation}
capturing the evolution of all matrices. The $P_{\nu}$ are the $4\times4$
equivalent of the previously defined matrices, see Eqs.~(\ref{eq:ansatz_mk4}).
Here the recursions for the parameters read
\begin{widetext}
\begin{equation}
\begin{aligned}a_{k+1} & =\frac{-8a_{k}+5a_{k}^{3}+a_{k}^{4}+\left(8+4a_{k}-22a_{k}^{2}-a_{k}^{3}+5a_{k}^{4}\right)b_{k}+\left(4+21a_{k}+3a_{k}^{2}-30a_{k}^{3}-4a_{k}^{4}\right)b_{k}^{2}+a_{k}\left(5+13a_{k}-4a_{k}^{2}-16a_{k}^{3}\right)b_{k}^{3}}{-16-4a_{k}+13a_{k}^{2}+5a_{k}^{3}+\left(-4-30a_{k}+3a_{k}^{2}+21a_{k}^{3}+4a_{k}^{4}\right)b_{k}+\left(5-a_{k}-22a_{k}^{2}+4a_{k}^{3}+8a_{k}^{4}\right)b_{k}^{2}+\left(1+5a_{k}-8a_{k}^{3}\right)b_{k}^{3}},\\
b_{k+1} & =\frac{-8b_{k}+b_{k}^{3}+a_{k}^{3}b_{k}\left(5+4b_{k}-16b_{k}^{2}\right)+a_{k}^{2}\left(4+13b_{k}-26b_{k}^{2}-12b_{k}^{3}\right)+a_{k}\left(8-12b_{k}-18b_{k}^{2}+b_{k}^{3}\right)}{-16-12a_{k}+a_{k}^{2}+a_{k}^{3}+2\left(2-13a_{k}-9a_{k}^{2}\right)b_{k}+\left(5+13a_{k}-12a_{k}^{2}-8a_{k}^{3}\right)b_{k}^{2}+4a_{k}\left(1+2a_{k}\right)b_{k}^{3}}
\end{aligned}
\label{eq:rg_recursions_mk4}
\end{equation}
\end{widetext}
with $a_0=b_0=z$ as the initial conditions. These recursions resemble those in Eqs. (\ref{eq:rg_recursions_mk3}), but the degrees of the polynomials in numerator and denominator are higher. This is a direct consequence of the higher number of sites eliminated during one iteration. Again, we have chosen a parametrization where $|a_{k+1}| = |b_{k+1}| = 1$ if $|a_k| = |b_k| = 1$. The rescaling of the phase of $a_k$ is shown in Fig. 5. The direct simulation for MK4 in Fig. 3 again confirms the RG prediction.

\paragraph*{RG for HN3:}

The derivation of RG equations for HN3, see Fig.~\ref{fig:rg_illustration_hn3},
are slightly more complicated than the above calculations for MK3
and MK4 for three reasons. First, the recursion on HN3 requires the
introduction of a fourth hopping parameter $D$ which is not present
in the actual graph, but becomes necessary to close the RG flow. Secondly,
the symmetry of the hoppings is not preserved by the recursions. This
means, after one decimation step, the matrix representing the hop
from 1 to 2 is no longer identical with the one from 2 to 1. Lastly,
the rules leading to HN3 inherently distinguish between even and odd
sites. As a result, the self-interaction terms become different for
those two groups. If we make the ansatz 
\begin{equation}
\begin{aligned}A & =\begin{bmatrix}\frac{b-a}{4} & \frac{a+b+2c}{4} & 0\\
0 & 0 & 0\\
0 & 0 & 0
\end{bmatrix}\cdot\mathcal{C}_{G} & C & =\begin{bmatrix}0 & 0 & 0\\
0 & 0 & 0\\
0 & 0 & z
\end{bmatrix}\cdot\mathcal{C}_{G}\\
B & =\begin{bmatrix}0 & 0 & 0\\
\frac{a+b+2c}{4} & \frac{b-a}{4} & 0\\
0 & 0 & 0
\end{bmatrix}\cdot\mathcal{C}_{G} & D & =\begin{bmatrix}0 & \frac{b-a}{4} & 0\\
0 & 0 & 0\\
0 & 0 & 0
\end{bmatrix}\cdot\mathcal{C}_{G}\\
M_{1} & =\begin{bmatrix}\frac{a+b-2c}{4} & \frac{b-a}{4} & 0\\
\frac{b-a}{4} & \frac{a+b-2c}{4} & 0\\
0 & 0 & 0
\end{bmatrix}\cdot\mathcal{C}_{G}\\
M_{2} & =\begin{bmatrix}\frac{a+b-2c}{4} & 0 & 0\\
0 & \frac{a+b-2c}{4} & 0\\
0 & 0 & 0
\end{bmatrix}\cdot\mathcal{C}_{G}
\end{aligned}
\label{eq:parametrization_hn3}
\end{equation}
we can take everything into account by writing the linear system corresponding
to the top right graphlet in Fig.~\ref{fig:rg_illustration_hn3}:
\begin{equation}
\begin{aligned}\tilde{\psi}_{4} & =A^{T}\tilde{\psi}_{1}+B^{T}\tilde{\psi}_{2}+M_{1}\tilde{\psi}_{4}+C\tilde{\psi}_{5}\\
\tilde{\psi}_{5} & =A^{T}\tilde{\psi}_{2}+B^{T}\tilde{\psi}_{3}+C^{T}\tilde{\psi}_{4}+M_{1}\tilde{\psi}_{5}
\end{aligned}
\label{eq:hn3_linear_system_1}
\end{equation}
Here $A^{T}$ represents the transpose of $A$. As it turns out, this
correctly describes the hopping in different direction (left or right
in the figure).

\begin{figure}
\includegraphics{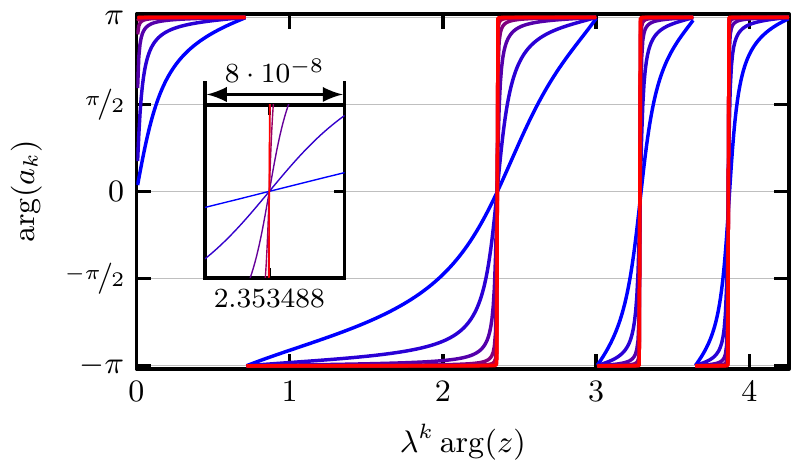} \protect\protect\caption{(Color online) Rescaling for MK4 of the phase of the first RG parameter
$a_{k}$ in Eq. (\ref{eq:rg_recursions_mk4}) around the fixed point
$z=1$ with $\lambda=\sqrt{\frac{247}{7}}$. The insets show a magnification
to illustrate the conversion towards a step function. In the main panel, $k = 2,6,...,14$ while $k = 20,22,...30$ for the inset. This corresponds to a system size of $N=13^{30}\approx 10^{34}$.}
\label{fig:mk4_rg_plot} 
\end{figure}

\begin{figure}
~\hfill{}\includegraphics{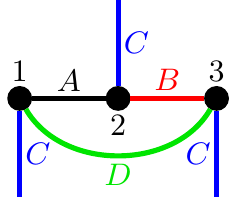} \hfill{}\includegraphics{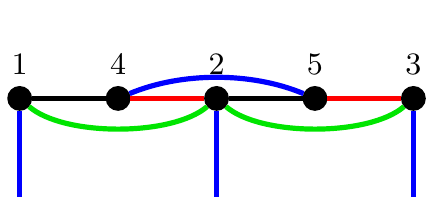}
\hfill{}~\\[8pt] ~\hfill{}\includegraphics{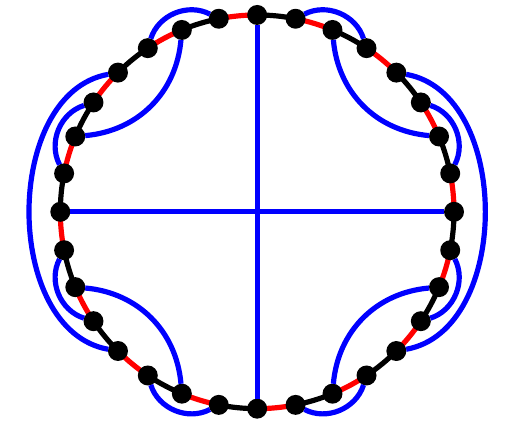}
\hfill{}~ \protect\caption{(Color online) Illustration of the decimation scheme for HN3. Growing
the network means inserting new nodes ($4$ and $5$) and connecting
them accordingly (top row). The graph at generation $k=5$ is shown
in the lower panel. The RG decimation requires an extra set of hopping
matrices ($D$, orange) in order to close the recursions, but these
are not present in the actual network.}
\label{fig:rg_illustration_hn3} 
\end{figure}

\begin{figure}
\includegraphics{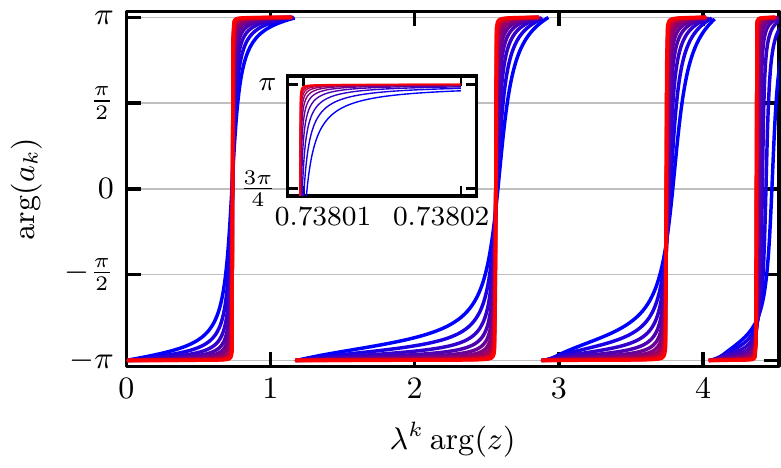} \protect\protect\caption{(Color online) Rescaling for HN3 of the phase of the first RG parameter
$a_{k}$ in Eq. (\ref{eq:hn3_recursions}) around the fixed point
$z=1$. The insets show a magnification to illustrate the conversion
towards a step function. In the main panel, $k=10,12,\dots,30$ while
$k=60,62,\dots80$ for the inset. This means the largest system size
is $N\approx10^{24}$. $\lambda=2^{1-\log_{2}(\varphi)/2}$, where
$\varphi=\left(\sqrt{5}+1\right)/2$ is the ``golden section'' .}
\label{fig:hn3_rg_plot} 
\end{figure}

\begin{figure}
~\hfill{}\includegraphics{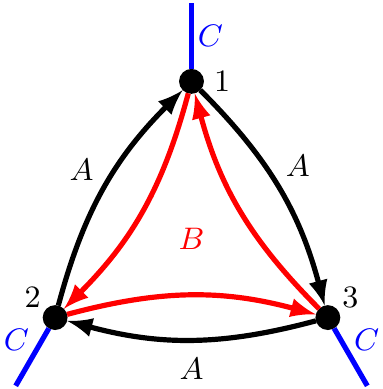} \hfill{}\includegraphics{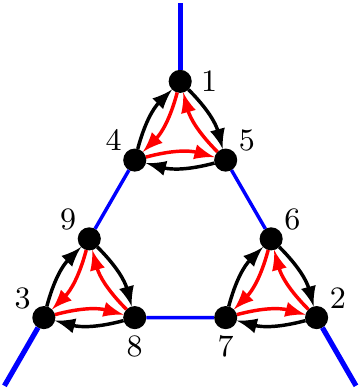}\hfill{}~\\[8pt]
~\hfill{}\includegraphics[width=0.9\linewidth]{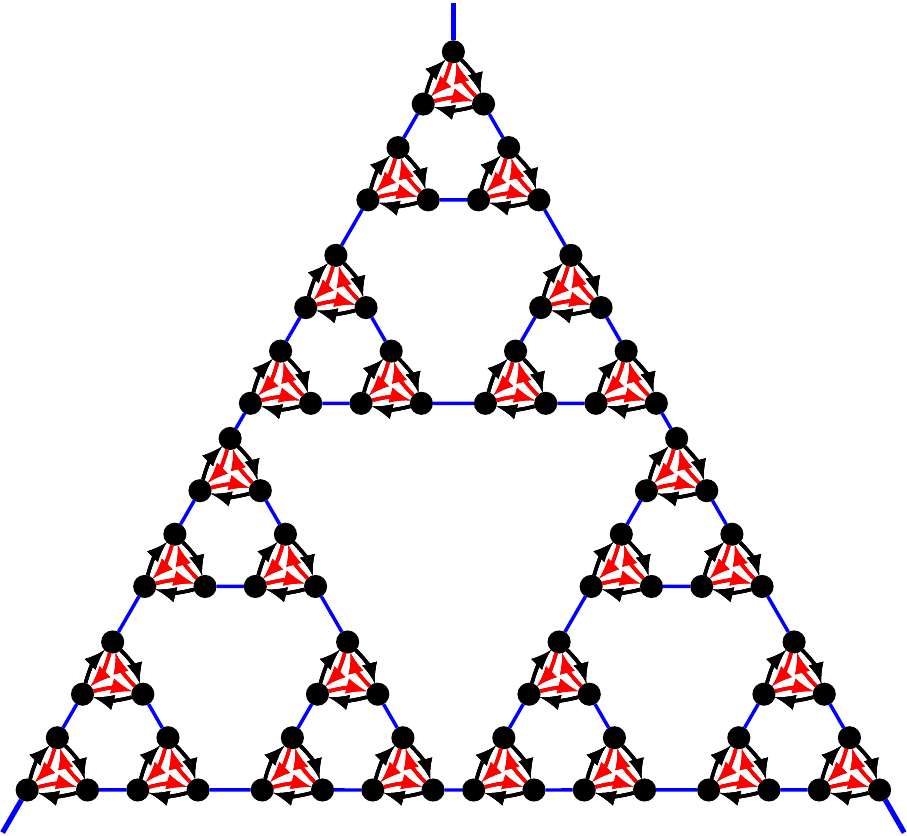}
\hfill{}~ \protect\caption{(Color online) The well know recursion generating the DSG (top row).
To make the positions of the hopping matrices also self-similar, we
have to introduce directionality of the hopping matrices $A$ and
$B$. The third one, $C$, is still symmetric. The lower panel shows
the system at generation four.}
\label{fig:dsg_rg_illustration} 
\end{figure}

By solving these equations for $\tilde{\psi}_{4}$ and $\tilde{\psi}_{5}$,
and inserting this into the equations for the remaining sites, 
\begin{equation}
\begin{aligned}\tilde{\psi}_{1} & =M_{2}\tilde{\psi}_{1}+D\tilde{\psi}_{2}+D^{T}\tilde{\psi}'_{2}+A\tilde{\psi}_{4}+B^{T}\tilde{\psi}_{5}'+C\tilde{\psi}_{*},\\
\tilde{\psi}_{2} & =D\tilde{\psi}_{1}+M_{2}\tilde{\psi}_{2}+D^{T}\tilde{\psi}_{3}+B^{T}\tilde{\psi}_{4}+A\tilde{\psi}_{5}+C\tilde{\psi}_{*},
\end{aligned}
\label{eq:hn3_linear_system_2}
\end{equation}
where we omitted the equation for $\tilde{\psi}_{3}$ as it is identical
to the first one. Every node is connected to a node of unknown index,
$\tilde{\psi}_{*}$, but the corresponding hopping matrix $C$ does
not change. After some algebra, we find the following recursion equations
for the three RG variables: 
\begin{equation}
\begin{aligned}a_{k+1} & =\frac{c_{k}(-3+z)-b_{k}(-3+z+c_{k}(-2+6z))}{6-b_{k}+c_{k}+(-2+3b_{k}-3c_{k})z}\,,\\
b_{k+1} & =\frac{c_{k}(3+z)-b_{k}(3+z+c_{k}(2+6z))}{-6+b_{k}-c_{k}+(-2+3b_{k}-3c_{k})z}\,,\\
c_{k+1} & =\frac{c_{k}+a_{k}(-1+2c_{k})}{2+a_{k}-c_{k}}\,,
\end{aligned}
\label{eq:hn3_recursions}
\end{equation}
with the initial conditions 
\begin{equation}
\begin{aligned}a_{0} & =\frac{z^{2}(1-3z)}{3-z}\,, & b_{0} & =\frac{z^{2}(1+3z)}{3+z}\,, & c_{0} & =z^{2}\,.\end{aligned}
\label{eq:hn3_initial_conditions}
\end{equation}
Again, we have chosen our Ansatz such that the variables stay of modulus
one when they start out that way. This time we show the rescaling
of the argument of the first RG parameter in Fig.~\ref{fig:hn3_rg_plot}.
As verification, we have also scaled the numerically obtained PDF
in Fig.~\ref{fig:direct_simulations}.

\paragraph*{RG for DSG:}

Finally, we consider the DSG again \cite{QWNComms13} with this approach,
see Fig.~\ref{fig:dsg_rg_illustration}. In order to make it renormalizable,
we have to introduce a directionality represented by the arrows for
$A$ and $B$. This just means that applying one hopping matrix, say
$A$, twice describes the hopping from site 1 to 2 (over 3), and not
1 to 3 back to 1. The matrix $C$ is not affected by this.

The linear system we need to solve in this case reads 
\begin{equation}
\begin{pmatrix}\tilde{\psi}_{4}\\
\tilde{\psi}_{5}\\
\tilde{\psi}_{6}\\
\tilde{\psi}_{7}\\
\tilde{\psi}_{8}\\
\tilde{\psi}_{9}
\end{pmatrix}=\begin{bmatrix}B & 0 & 0 & M & A & 0 & 0 & 0 & C\\
A & 0 & 0 & B & M & C & 0 & 0 & 0\\
0 & B & 0 & 0 & C & M & A & 0 & C\\
0 & A & 0 & 0 & 0 & B & M & C & 0\\
0 & 0 & B & 0 & 0 & 0 & C & M & A\\
0 & 0 & A & C & 0 & 0 & 0 & B & M
\end{bmatrix}\begin{pmatrix}\tilde{\psi}_{1}\\
\tilde{\psi}_{2}\\
\tilde{\psi}_{3}\\
\tilde{\psi}_{4}\\
\tilde{\psi}_{5}\\
\tilde{\psi}_{6}\\
\tilde{\psi}_{7}\\
\tilde{\psi}_{8}\\
\tilde{\psi}_{9}
\end{pmatrix}\label{eq:dsg_linear_system1}
\end{equation}
The results then has to be plugged into the equations for $\tilde{\psi}_{1},\dots,\tilde{\psi}_{3}$:
\begin{equation}
\begin{aligned}\tilde{\psi}_{1} & =A\tilde{\psi}_{4}+B\tilde{\psi}_{5}+C\tilde{\psi}'_{2,3}\\
\tilde{\psi}_{2} & =A\tilde{\psi}_{6}+B\tilde{\psi}_{7}+C\tilde{\psi}''_{1}\\
\tilde{\psi}_{1} & =A\tilde{\psi}_{8}+B\tilde{\psi}_{9}+C\tilde{\psi}'''_{1}
\end{aligned}
\label{eq:dsg_linear_system2}
\end{equation}
Here the algebra is very involved, and we have shown elsewhere \cite{QWNComms13}
how it can done. There, we showed the scaling of the parameters and
deduced $d_{w}^{QW}$ from it using the RG. The scaling plot obtained
by direct simulations in Fig.~\ref{fig:direct_simulations} confirms
again the conjecture.

\bibliographystyle{apsrev4-1}
\bibliography{/Users/stb/Boettcher}

\end{document}